\documentclass[12pt]{iopart}
\usepackage{graphicx}
\usepackage{amssymb}
\usepackage{bm}
\usepackage{color}

\begin{document}

\title{Universality of the triplet contact process with diffusion}
\author{R D Schram$^1$ and G T Barkema$^{1,2}$}
\address{$^1$ Instituut-Lorentz, Leiden University, 
  P.O. Box 9506, 2300 RA  Leiden, The Netherlands}
\address{$^2$ Institute for Theoretical Physics, Utrecht University, 
  P.O. Box 80195, 3508 TD  Utrecht, The Netherlands}
\eads{\mailto{schram@lorentz.leidenuniv.nl} \mailto{g.t.barkema@uu.nl}}

\begin{abstract}
The one-dimensional triplet contact process with diffusion (TCPD) model
has been studied using fast multispin GPU Monte Carlo simulations. In
particular, the particle density $\rho$ and the density of pairs of
neighboring particles $\rho_p$ have been monitored as a function of
time. Mean field predictions for the time evolution of these observables
in the critical point are $\rho\sim t^{-\delta}$ and $\rho_p\sim
t^{-\delta_p}$ with $\delta=1/3$ and $\delta_p=2/3$. We observe that in
the vicinity of the critical point of the model, the ratio $\rho_p/\rho$
tends to a constant, which shows that the one-dimensional TCPD model is
not described by mean field behavior.  Furthermore, our long simulations
allow us to conclude that the mean field prediction of the exponent
$\delta$ is almost certainly not correct either. Since the crossover to the
critical regime is extremely slow for the TCPD model, we are unable
to pinpoint a precise value for $\delta$, though we find as an upper
bound $\delta < 0.32$. 
\end{abstract}
\pacs{05.10.Ln, 05.50.+q, 64.60.Ht}

\noindent{\it Keywords\/}: TCPD, PCPD, dynamic universality, reaction-diffusion models
\maketitle

\section{Introduction}

The triplet contact process with diffusion (TCPD) model belongs to a
set of closely related models of ``fermionic'' particles on a lattice that
follow very simple dynamical rules. With {\it fermionic} we mean in
this context that only one particle can be present at each site. These
models have been studied extensively, because they were conjectured
by Grassberger \cite{grass} and Janssen \cite{janssen} to belong to a
relatively small number of dynamical universality classes, determined by
coarse features such as the dimensionality, symmetries in the model,
and conservation laws; in close analogy to universality classes in
equilibrium statistical physics.

Historically, the first of these models to be studied extensively is the
Directed Percolation (DP) model. This model lends itself extremely well
for computer simulations, and its exponents are therefore known with high
accuracy.  For example, using exact enumeration techniques, the exponent
$\delta$ was found to be 0.159464(6) \cite{jensen99}, where $\delta$ is
defined through the time dependence of the particle density $\rho$ of the system,
starting from a state with a uniform high density:

\begin{equation}\label{eq:rho}
\rho \sim t^{-\delta}.
\end{equation}

Even though this model is very easy from a numerical point of view,
there are no theoretical predictions for these exponents, not even in
one dimension.  Since the Grassberger--Janssen conjecture says that the
DP model belongs to a larger class of models with the same critical
exponents, much effort has been undertaken to verify numerically the
exponents in models that are also expected to be in the same universality
class.

In one interpretation of the DP model, particles are placed on a
lattice, and follow two reactions: with a statistical rate $p$, each
particle annihilates, and with a statistical rate $1-p$, each particle
creates a new particle on an adjacent site, provided that it is
vacant. Sometimes, the particles can also hop to neighboring lattice
sites with a diffusion rate $d$. An extension of the DP model that
has been studied thoroughly in literature is the pair contact process
with diffusion (PCPD) model, in which the annihilation and procreation
reactions can only take place if two particles are placed next to each
other (and with $d>0$).  There has been a lot of discussion about the
critical exponents, mainly $\delta$, with estimates ranging from the
DP-value $\delta=0.159$ \cite{small,barkema03} to $\delta=0.35$ \cite{nohpark}.
In Ref.~\cite{schram_pcpd}, we performed simulations of the PCPD model in which the triple product
$V=N\times L \times T$, where $N$ independent simulations are performed
with a lattice of size $L$ over a time range $T$, is much larger than in
previous studies. The resulting data showed that in the case of the PCPD
model, finite-time corrections are particularly severe, and with careful
analysis we found more evidence suggesting a value close to the DP value,
than the contrary.

Here, we simulate the TCPD model, which makes annihilation and
procreation conditional on triplets of particles.  Currently, the claim
in literature is that TCPD is fundamentally different from both the
PCPD and DP models, as evidenced by different values for exponents, e.g. $\delta=0.27(1)$\cite{kockelkoren}
and $\delta=1/3$\cite{odor_tcpd,vdburg}.  We use the same efficient algorithm,
described in \cite{schram_gpu}, to reach a triple product ($N\times L
\times T$) which is almost three orders of magnitude larger than in these previous
studies.  We find that for shorter times, the exponent $\delta$ is very
close to its mean field prediction of $1/3$, but for very long times, this
exponent starts to drift. We attribute this to finite-time effects that
are even stronger than in the case of PCPD. Even with our fast algorithm
and long simulations, we were not able to reach time scales that make
it possible to retrieve an accurate estimate for $\delta$. Importantly,
however, we find that the DP value cannot be excluded, meaning that it
is not yet proven that the TCPD model falls outside the DP universality class.

\section{Method}

Several slightly different versions of the TCPD model have been
studied. We use a version with the reactions $3A \rightarrow 4A$ and $3A \rightarrow
\emptyset$, which is described by the following reactions and rates:

\begin{eqnarray}
\begin{array}{ccc}
\left\{
\begin{array}{ccc}
AAA0 & \rightarrow & AAAA \\
0AAA & \rightarrow & AAAA
\end{array} \right.
&{\rm each\,\,with\,\, rate}&
(1-p)(1-d)/2 \label{2Ato3A} \\
AAA~ \rightarrow ~ 000
&{\rm with \,\,rate}& p\,(1-d) \label{2Ato0}\\
A0  \leftrightarrow  0A &{\rm with \,\,rate}&  d \label{A0to0A}
\end{array}
\label{eq:rates}
\end{eqnarray}

 In a straightforward implementation of this model, first the type of reaction
is selected, based on a random number $r \in \left[0,1\right>$, and depending
on its value, one of the reactions is proposed (but not always carried out):

\begin{itemize}
\item if $r<d$, a random pair of neighboring sites $\{i,i+1\}$ is selected; in case one site is occupied and the
other is empty, the particle hops from the occupied site to the empty one.
\item else if $r-d<p(1-d)$, a random triplet of sites $\{ i, i+1, i+2 \}$ is
selected; if all three sites are occupied, they are all made vacant.
\item else if $r-d-p(1-d)<(1-p)(1-d)/2$,  a random triplet of sites $\{ i, i+1, i+2\}$ is
selected; if all three sites are occupied and site $i+3$ is vacant, a particle is placed on this last site.
\item else,  a random triplet of sites $\{ i, i+1, i+2\}$ is
selected; if all three sites are occupied and site $i-1$ is vacant, a particle is placed on this last site.
\end{itemize}
Irrespective the type of reaction and its success, the time scale is incremented by $\Delta t=1/N$.
These steps are then iterated many times.
As noted in the introduction, for the results presented here, we used as a basis an algorithm that
leverages the power of graphics processing units (GPUs), which is
described in detail in \cite{schram_gpu}.

Before one can actually measure the exponent $\delta$, a prerequisite is
to determine the critical
annihilation rate $p_c$. For low annihilation rates ($p<p_c$), the system
will, given enough time, settle with extremely high probability for a more or less
constant particle density. This regime is called the active regime. On the other
hand, in the inactive regime with high annihilation rates ($p>p_c$), the
particles will quickly die out. In between, exactly at $p=p_c$, there is the
critical regime, where the density decreases slower than in the inactive
regime, following a power-law decay with a critical exponent $\delta$.

Thus, it is necessary to first identify an estimate for the critical point $p_c$, before we
can estimate the critical exponent $\delta$. To this
effect we use as our main tool the effective exponent $\delta_{\rm eff}$
as a function of time, defined as:

\begin{equation}\label{eq:numdif}
\delta_{\rm eff}(t=\sqrt{t_1 t_2}) = -\frac{\log(\rho(t_1))-\log(\rho(t_2))}{\log(t_1)-\log(t_2)},
\end{equation}

and a similar expression for the effective exponent $\delta_{{\rm eff},p}$ for the pair density.
Substituting the asymptotic behavior as given in eq.~(\ref{eq:rho}),
we retrieve that $\delta_{\rm eff} \rightarrow \delta$ as $t$ goes to
infinity. The procedure is equivalent to numerical differentiation of $\partial \log(\rho)/\partial \log(t)$.
Equation (\ref{eq:numdif})
shows that there is still freedom in choosing $t_1$ and $t_2$.
The trade-off is as follows: if we choose $t_1$ closer to $t_2$, the plot
is generally more accurate, in the sense that features present in the analytical
$\delta_{\rm eff}$ curve are less smoothed out and lost that way. On
the other hand, the curve is much more noisy.
We found that in our case choosing $t_1/t_2 \approx \exp(3)$ gives good results.

We find an estimate for $p_c$ by a manual binary search, which sounds more
cumbersome than it is, because far from $p_c$ the $\delta_{\rm eff}$ plots
are very clearly recognizable as either sub- or super-critical. Closer to
$p_c$ it gets much harder to estimate $p_c$, because of the drift in the
effective exponent, which gives our estimate for $p_c$ a larger error bar.
Apart from the density $\rho$ of the system, we can also measure the pair
density $\rho_p$. This is especially useful for testing
the validity of mean field theory, as it predicts that $\rho_p = \rho^2$,
and thus also $\delta_{{\rm eff},p} = 2 \delta_{\rm eff}$.

For each value of $p$ we performed at least $N=200$ independent
simulations of runs up to $T=8 \cdot 10^8$, with a lattice size
of $L=2^{21} = 2097152$. We simulated at annihilation rates between
$p=0.09500$ and $p=0.09513$: 0.095, 0.09504, 0.09508, 0.09511, 0.09512,
0.095125, 0.09513, always with $d=0.5$. Our estimate for the critical annihilation rate
is $p_c=0.09510_{-0.00005}^{+0.00002}$.

\section{Simulation results}

\begin{figure}\centering
\includegraphics[width=10cm]{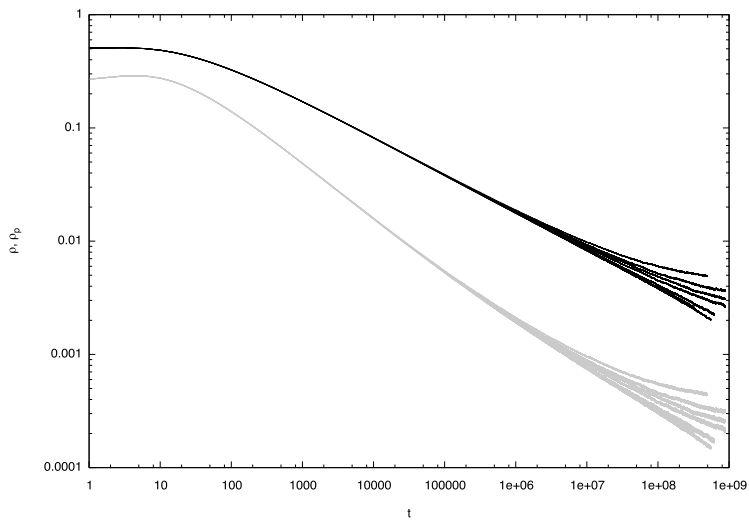}
\caption{The density $\rho$ (black curves) and the pair density $\rho_p$ (gray curves) as a function of time in a double logarithmic plot. The lattice size is $L=2^{21}$, with 200 independent runs for each value of $p$ and $0.095<p<0.09513$.}
\label{fig:rho}
\end{figure}

First, in figure \ref{fig:rho} we present the raw simulation data of the average particle density $\rho$ and the pair density $\rho_p$ as a function of time, starting from a random initial state with density $\rho = 1/2$ (and $\rho_p=1/4$). The particle density shows an approximately straight line in a double logarithmic plot, indicating power-law behavior. At the same $p$ values, the pair density shows strong curvature, which indicates strong finite-time corrections to power-law behavior. We note that curves close to the critical point look qualitatively similar to the ones found in Ref. \cite{kockelkoren}, although the slope is different: we find $0.33$, whereas they reported $0.27(1)$. In our opinion, this discrepancy can be attributed to differences in finite-time corrections, due to details in the models. Because our data has very small error bars, we can use $\delta_{\rm eff}$ to investigate the ``power-law'' like behavior of the density and the pair density in more detail.

\begin{figure}\centering
\includegraphics[width=10cm]{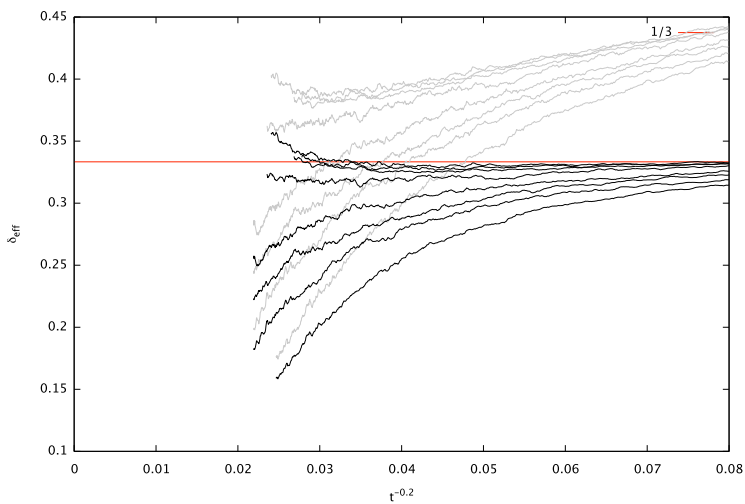}
\caption{The effective exponent $\delta_{\rm eff}$ as obtained for both
the particle density $\rho$ (colored black) and the pair density $\rho_p$
(gray), as a function of $t^{-0.2}$. The lattice size is $L=2^{21}$, with 200 independent runs for each value of $p$ and $0.095<p<0.09513$.}
\label{fig:delta_eff}
\end{figure}

The effective exponents $\delta_{\rm eff}$ and $\delta_{{\rm eff},p}$, describing the decay
of particles and of pairs of particles, are plotted against $t^{-\alpha}$ in
figure \ref{fig:delta_eff}, with values for $p$ close to $p_c$.  We chose $\alpha=0.2$, as this
is a rough estimate for the exponent governing the leading finite-time corrections.
As we will not use any extrapolation method, the choice for $\alpha$ does not affect
our conclusions. 

It is clear from the plot that $\delta_{{\rm eff},p} \neq 2 \delta_{\rm
eff}$, even at short times $t$, and thus we conclude that mean field
theory does not correctly describe the TCPD model.

The figure also shows that some curves first ascend to a value
for $\delta_{\rm eff}$ of mean field theory ($1/3$), then turn slightly
downwards to values of around $0.32$, and then ascend above $1/3$
again. We interpret this behavior as a signature of a $p$-value which
is close to, but slightly below $p_c$, as the data roughly follow
the critical curve, before reaching the inactive regime. Our main
reason is that we find it very unlikely that the critical curve makes
more than one bend on these time scales. Thus, we find a lower bound:
$\delta<0.32$. Obviously, if our assumption is not correct this lower
bound is also not correct. However, were this the case we believe that
any attempt at numerical analysis will be impossible with the current
simulation approach and state of software and hardware. Another important
aspect to note is that $\delta_{\rm eff}$ and $\delta_{{\rm eff},p}$
are closing in on each other in a way very similar to that in the PCPD
model \cite{schram_pcpd}.

\begin{figure}\centering
\includegraphics[width=10cm]{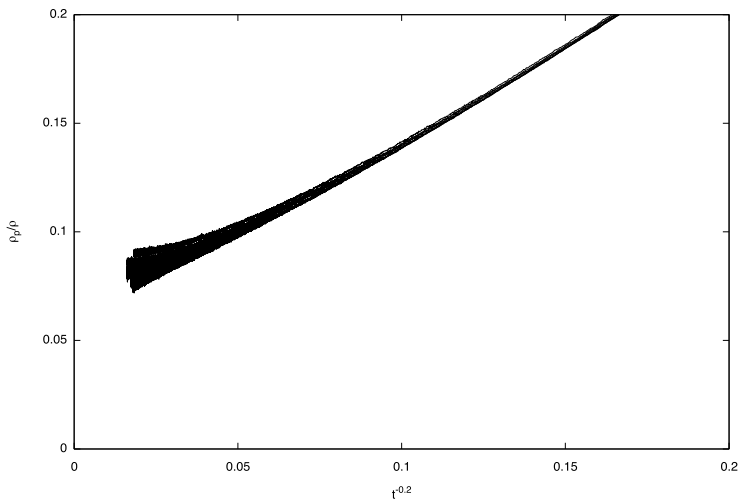}
\caption{The ratio $\rho_p/\rho$ between the pair density $\rho_p$ and the particle
density $\rho$ as a function of $t^{-\alpha}$, with $\alpha=0.2$. The precise value of $\alpha$
is not important here. At criticality the curve seems to arrive at the
vertical axis at a value higher than $0$, which means that eventually
$\rho/\rho_p$ will reach a constant value.}
\label{fig:ratio}
\end{figure}

The ratio $\rho_p/\rho$ is plotted in figure \ref{fig:ratio} against
$t^{-\alpha}$, again with $\alpha=0.2$, for different values of $p$,
including ones that are above $p_c$. We have to be more careful here
in our choice of $\alpha$, because we are trying to make a conclusion
about the extrapolated value. If we assume that $\rho_p/\rho \sim
t^{-\beta}$, then the choice $\alpha=\beta$ combined with linear extrapolation
yields the correct constant for $t \rightarrow \infty$. On
the other hand, choosing $\alpha>\beta$ yields an overestimation,
whereas $\alpha<\beta$ gives us a prediction lower than the true
value. Since we are not interested in the exact value of the ratio at
$t\rightarrow \infty$, but only in whether it is equal to 0 or higher,
we try to choose $\alpha<\beta$. Since $\beta$ is unknown, we chose
a value for $\alpha$, such that the critical curves seem to arrive close to
horizontally at the x-axis, which is indicative of $\alpha \lesssim \beta$. In
figure \ref{fig:ratio}, all curves are approaching the vertical axis
(corresponding to $t\rightarrow \infty$) in a way that strongly suggests
that it will go to a finite value. Thus, we conclude that the ratio
$\rho_p/\rho$ approaches a non-zero value as $t\rightarrow \infty$.

\section{Summary and conclusion}

We have performed extensive simulations of the one-dimensional TCPD
model, using a highly efficient GPU-based simulation approach. We find
that the TCPD model is not described by mean field theory, as evidenced
by the convergence of the ratio of the pair density and the particle
density $\rho_p/\rho$ to a non-zero value at criticality, instead of
going to zero as expected by mean field theory. Given the similarities
between the PCPD model and the TCPD model with regards to finite-time
corrections, we find it not unlikely that they both belong to the DP
universality class. Numerical evidence for the PCPD model belonging to
the DP universality class was given by us in Ref. \cite{schram_pcpd}. We
emphasize that there is no solid numerical evidence propositioning
a value of $\delta$ for the TCPD model. However, we do find an upper
bound $\delta<0.32$, which excludes most previous literature values for
this exponent in TCPD, and is also less than mean field. Thus a safer
conclusion is that numerical data does not exclude the possibility that
TCPD and PCPD belong to the DP universality class, and that this should
still be considered as a serious possibility.

\ack
Computing time on the ``Little Green Machine", which is funded by the Dutch
agency NWO, is acknowledged. We thank Gertjan van den Burg for useful discussions.

\end{document}